\documentclass[a4paper]{jpconf}
\usepackage{graphicx}
\usepackage[reqno]{amsmath}
\usepackage{amssymb}

\newcommand{\bea}{\begin{equation}\begin{array}{c}}
\newcommand{\eea}{\end{array}\end{equation}}
\newcommand{\ea}{\end{array}} 
\newcommand{\beq}{\begin{equation}}
\newcommand{\eeq}{\end{equation}}
\newcommand{\bad}{\begin{array}{ccc}}

\newcommand{\mefff}{\mbox{$ < \! m \! > $}}
\newcommand{\ba}{\begin{array}{c}}

\newcommand{\half}{\frac{1}{2}}
\newcommand{\diag}{{\rm diag}}

\newcommand{\betabeta}{\mbox{$(\beta \beta)_{0 \nu}  $}}
\newcommand{\meff}{\mbox{$\left|  < \!  m \!  > \right| $}}

\begin{document}

\title{Indirect Signatures of Type I See-Saw Scenarios}

\author{Emiliano Molinaro}
\address{ Centro de F\'{i}sica Te\'{o}rica de Part\'{i}culas (CFTP), 
        Instituto Superior T\'{e}cnico, 
         Technical University of Lisbon, 1049-001, Lisboa, Portugal}
\ead{emiliano.molinaro@ist.utl.pt}

\begin{abstract}
We consider the low energy constraints that can be applied to type I see-saw extensions of the Standard Model
in which the right-handed neutrinos are taken at the electroweak scale.  In the reported scenarios, 
the flavour structure of the charged current and neutral current weak interactions of the Standard Model leptons with 
the heavy right-handed neutrinos is essentially determined by the neutrino oscillation parameters. In this case,
correlations among different measurable phenomena in the lepton sector 
may provide compelling indirect evidence of low energy  see-saw mechanism of 
neutrino mass generation.
\end{abstract}

\section{Introduction}

The measurement of the solar and atmospheric  neutrino oscillation paramaters has provided compelling evidence 
for physics beyond the Standard Model (SM) of elementary particles.
Massive active neutrinos can be naturally accounted for in see-saw type extensions of the SM, where new 
fermion and/or scalar representations are introduced in the theory with suitable Yukawa couplings to the SM lepton doublets
 \cite{seesaw}. 
The mass of the new physical states is in general unrelated to the electroweak (EW) symmetry breaking scale and, therefore, can
assume arbitrary large values up to the Planck scale.


On the purely phenomenological side, it is interesting to study see-saw scenarios
in which new physics is manifest at the TeV scale and can be probed in collider searches, LHC included. 
In this physical context, the phenomenology of type I see-saw extensions has been studied in detail in \cite{Ibarra:2010xw,Ibarra:2011xn}, in a model independent way. 
The new particle states in such scenarios consist of at least two heavy SM-singlet fermions, which  
are conventionally denoted as  right-handed (RH) neutrinos, $\nu_{aR}$  ($a>2$), and give rise, when EW symmetry
is broken, to the following mass terms in the Lagrangian
\begin{equation}
\mathcal{L}_{\nu}\;=\; -\, \overline{\nu_{\ell L}}\,(M_{D})_{\ell a}\, \nu_{aR} - 
\half\, \overline{\nu^{C}_{aL}}\,(M_{N})_{ab}\,\nu_{bR}\;+\;{\rm h.c.}\,, 
\label{typeI}
\end{equation}
%
where $\nu^{C}_{aL}\equiv C \overline{\nu_{aR}}^T$ ($a=1,2,\ldots,K$),
 $M_{N} = (M_{N})^T$ is the 
$K\times K$ Majorana mass matrix of the RH neutrinos
and $M_{D}$ provides the $3\times K$ neutrino Dirac 
mass term.
The Majorana mass $m_{\nu}$ for 
the active left-handed neutrinos is given by the well known see-saw relation:
$m_{\nu}\cong -  M_{D} M_{N}^{-1} (M_{D})^T$. 
After the diagonalization of the full mass matrix given in (\ref{typeI}), the charged
current (CC) and neutral current (NC) weak interactions involving the heavy
Majorana mass eigenstates $N_{j}$ ($j=1,2,\ldots,K$) can be expressed as \cite{Ibarra:2010xw}:
\begin{eqnarray}
 \mathcal{L}_{CC}^N &=& -\,\frac{g}{2\sqrt{2}}\, 
\bar{\ell}\,\gamma_{\alpha}\,(RV)_{\ell k}(1 - \gamma_5)\,N_{k}\,W^{\alpha}\;
+\; {\rm h.c.}\,\label{NCC},\\
 \mathcal{L}_{NC}^N &=& -\frac{g}{2 c_{w}}\,
\overline{\nu_{\ell L}}\,\gamma_{\alpha}\,(RV)_{\ell k}\,N_{k L}\,Z^{\alpha}\;
+\; {\rm h.c.}\;,\label{NNC}
\end{eqnarray}
with $R^{*}\cong M_{D}M_{N}^{-1}$ at leading order in the see-saw expansion
and $V^{T}M_{N}V\cong \diag(M_{1},M_{2},\ldots,M_{K})$.
The couplings $|(RV)_{\ell j}|$ can in principle be sizable, typically $ |(RV)_{\ell j}|\sim 10^{-(3\div2)}$  for $M_k\approx (100\div 1000)$ GeV. Then, in order to reproduce  small neutrino masses via the see-saw mechanism, 
a ``large'' contribution to $m_{\nu}$ from $N_{1}$ is \emph{exactly} cancelled by 
a negative contribution from a second RH neutrino, say $N_{2}$, provided:
\begin{equation}
(RV)_{\ell 2}=\pm i\, (RV)_{\ell 1}\sqrt{\frac{M_1}{M_2}}\,,
\label{rel0}
\end{equation}
where $M_{1,2}$ is the mass of the RH neutrino $N_{1,2}$. Barring accidental cancellations, 
relation (\ref{rel0}) is naturally fulfilled in models where an approximately conserved lepton charged exists. 
In such scenarios \emph{$N_{1}$ and $N_{2}$ form a pseudo-Dirac pair and the neutrino oscillation parameters fix the flavour structure of their weak CC and NC couplings to gauge bosons and charged leptons,
up to an overall constant $y$}  (see \cite{Ibarra:2010xw,Ibarra:2011xn} for a details):
\begin{eqnarray}
\label{mixing-vs-y}
\left|\left(RV\right)_{\ell 1} \right|^{2}&=&
\frac{1}{2}\frac{y^{2} v^{2}}{M_{1}^{2}}\frac{m_{3}}{m_{2}+m_{3}}
	\left|U_{\ell 3}+i\sqrt{m_{2}/m_{3}}\,U_{\ell 2} \right|^{2}\,,
~~{\rm NH}\,,\\
\left|\left(RV\right)_{\ell 1} \right|^{2}&=&
\frac{1}{2}\frac{y^{2} v^{2}}{M_{1}^{2}}\frac{m_{2}}{m_{1}+m_{2}}
	\left|U_{\ell 2}+i\sqrt{m_{1}/m_{2}}\,U_{\ell 1} \right|^{2}
\cong \;\frac{1}{4}\frac{y^{2} v^{2}}{M_{1}^{2}}
\left|U_{\ell 2}+iU_{\ell 1} \right|^{2}\,,
\,{\rm IH}\,,
\label{mixing-vs-yIH}
\end{eqnarray}
where $m_1$, $m_2$ and $m_3$ are the light  active neutrino masses in the case of 
a normal/inverted hierarchical (NH/IH) mass spectrum.

\section{Neutrinoless double beta decay in low scale see-saw scenarios}
 
The mass splitting of the two RH neutrinos is highly constrained from the experimental upper
limits set in neutrinoless double beta (\betabeta-) decay experiments. Indeed, in this case the effective
Majorana mass $\meff$, which controls the \betabeta-decay rate, receives an additional 
contribution from the exchange of the heavy Majorana neutrinos $N_{k}$, which may be
sizable/dominant for ``large'' couplings $(RV)_{\ell j}$. 
For $K=2$, given a nucleus $(A,Z)$, one has (see \cite{Ibarra:2010xw,Ibarra:2011xn} for details):
\begin{equation} 
\meff \cong 
\left |\sum_{i=1}^{3}U^2_{ei}\, m_i 
- \sum_{k=1}^{2}\, F(A,M_k)\, (RV)^2_{e k}\,M_k \right |\,,
\label{mee1}
\end{equation}
%
where for  $M_{k}=(100\div1000)$ GeV: 
$F(A,M_k)\cong(M_{a}/M_{k})^{2}f(A)$, 
$M_{a}\approx 0.9$ GeV and  $f(A)\approx 10^{-(2\div1)}$.
Using eq.~(\ref{rel0}), the $N_{k}$ contribution 
to the effective Majorana mass is simply
\begin{equation}
\mefff^{{\rm N}} 
\cong - \,\frac{2z + z^2}{(1 + z)^2}\,
\left(RV\right)_{e1}^{2}\, \frac{M_{a}^{2}}{M_{1}}\,f(A)\,,
\label{mee3}
\end{equation}
with $z\equiv |M_{2}-M_{1}|/M_{1}$.
In the case of \emph{sizable} couplings of RH neutrinos to the charged leptons, $e.g.$ $|(RV)_{\ell 1}|\approx 10^{-2}$, 
this contribution can be even as large as $|\mefff^N|\sim 0.2~(0.3)$ eV for $z\cong 10^{-3}\,(10^{-2})$
and $M_{1}\cong 100\, (1000)$ GeV \cite{Ibarra:2010xw,Ibarra:2011xn}.~\footnote{Therefore, in this scenario the two RH neutrinos $N_{1}$ and $N_{2}$ form a pseudo-Dirac pair. Notice that this conclusion is valid even in the case in which there is no conserved lepton charge in the limit of zero splitting at tree level between the masses of the pair \cite{Ibarra:2010xw,Ibarra:2011xn}.}~An effective Majorana mass of this order of magnitude may take place in both types of neutrino mass spectrum
and can be accessible in ongoing experiments looking for \betabeta-decay  ($e.g.$ the GERDA experiment \cite{GERDA},
which can probe values of $\meff$ up to $\sim 0.03$ eV). 

\section{Charged lepton radiative decays in low scale see-saw scenarios}

\begin{figure}
\begin{center}
\begin{tabular}{cc}
\includegraphics[width=7.5cm,height=6.5cm]{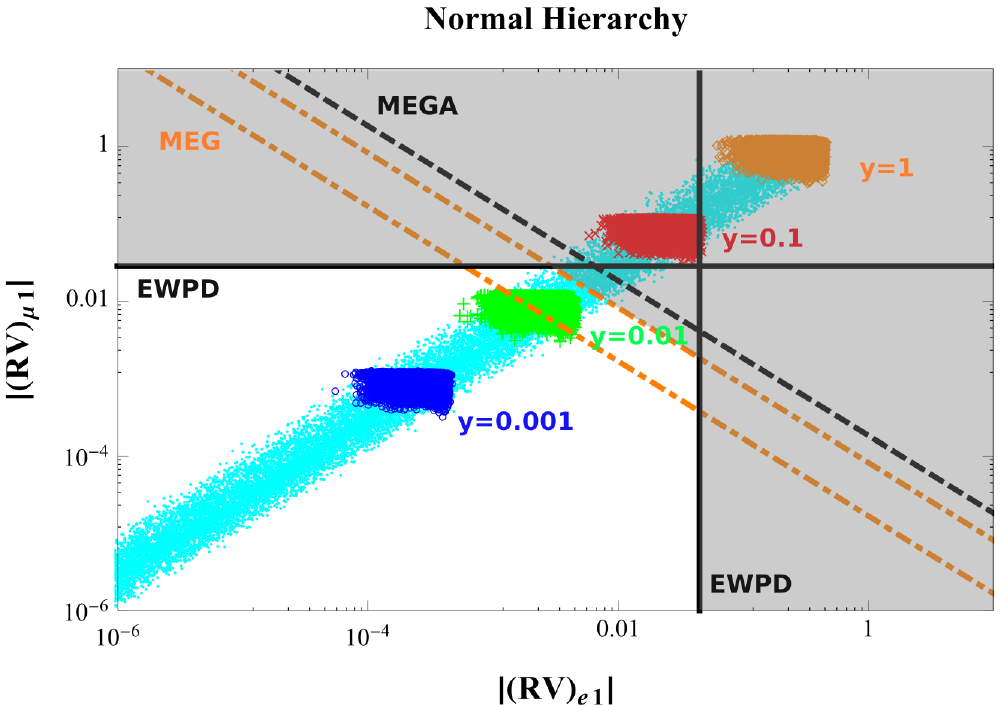} &
\includegraphics[width=7.5cm,height=6.5cm]{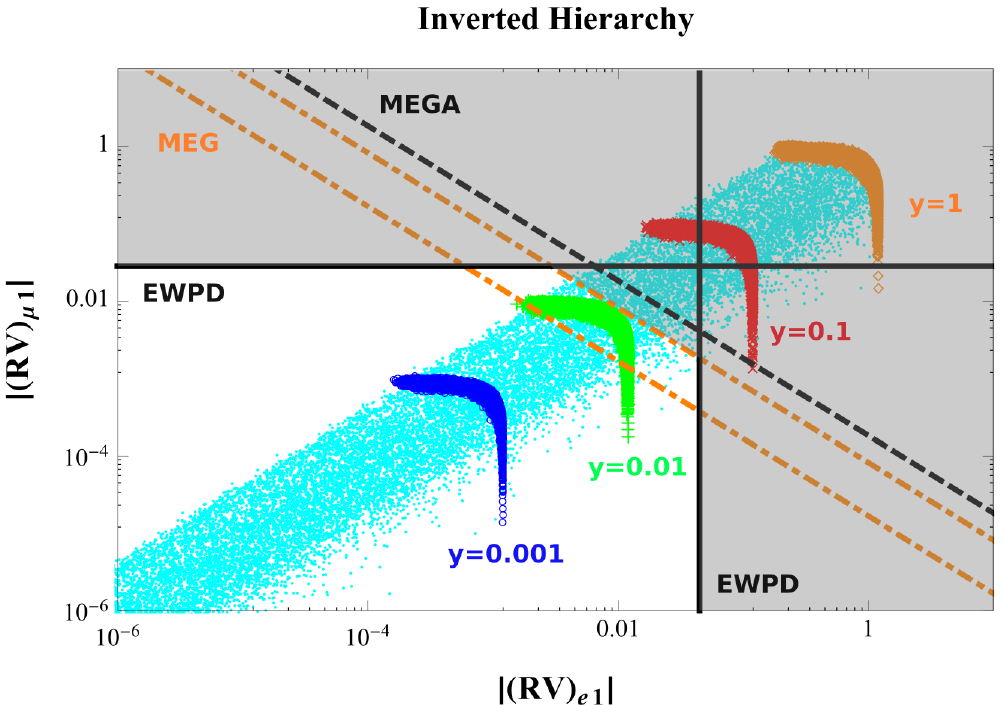}\\
\end{tabular}
\caption{Correlation between $|(RV)_{e1}|$ and $|(RV)_{\mu1}|$ in the case of NH (left panel)
and IH (right panel) light neutrino mass spectrum, for 
$M_{1}=100$ GeV and fixed values of $y$. 
The cyan points correspond
to random values of $y\leq 1$. \label{fig2}}
\end{center}
\end{figure}

In the scenario under discussion, lepton flavour radiative decays allow  to put further constraints on the size of the 
mixing between light and heavy Majorana neutrinos. The strongest bounds are obtained
from the current  upper limit on $\mu\to e+\gamma$ branching ratio \cite{Ibarra:2011xn}:
\begin{eqnarray}
&&B(\mu\to e+\gamma) =
\frac{\Gamma(\mu\to e+\gamma)}{\Gamma(\mu\to e+\nu_{\mu}+\overline{\nu}_{e})} 
=
\frac{3\alpha_{\rm em}}{32\pi}\,|T|^{2}\,,
\label{Bmutoeg1}\\
&&T=\sum\limits_{j=1}^{3}
\left[\left(1+\eta\right)U\right]_{\mu j}^{*} \,\left[\left(1+\eta\right)U \right]_{e j} G\left(\frac{m_{j}^{2}}{M_{W}^{2}}\right)
+ \sum\limits_{k=1}^{2} \left( RV\right)_{\mu k}^{*} \left( RV\right)_{e k} G\left(\frac{M_{k}^{2}}{M_{W}^{2}}\right)\nonumber\\
&&\;\;\;\;\cong\;2\left [(RV)_{\mu 1}^{*}\, (RV)_{e1}\right ] \left[ G(M_1^2/M_W^2) - G(0)\right]\,,
\label{T2}
\end{eqnarray}
where $\eta\equiv-RR^{\dagger}/2 $. The last relation arises from (\ref{rel0}) and taking into account that  $z\ll1$, 
because of $\meff$ upper limit. Therefore, taking $B(\mu\to e+\gamma)<2.4\times 10^{-12}$ at 90\% C.L.  from MEG
  experiment \cite{MEG}, 
the following constraint for $M_1 = 100~{\rm GeV}$ ($M_1 = 1$ TeV) is derived \cite{Ibarra:2011xn}
\begin{equation}
\left |(RV)_{\mu 1}^{*}\, (RV)_{e1}\right| < 0.8\times 10^{-4}\,
(0.3\times 10^{-4})\,.
\label{T3}
\end{equation}
This can be recast as an upper bound on the 
neutrino Yukawa parameter $y$ (see figure~\ref{fig2}) \cite{Ibarra:2011xn}:
\begin{eqnarray}
&& y\lesssim 0.036~(0.09)\,~{\rm for~NH~with~} M_1=100\,{\rm GeV}\,~{\rm and~}\sin\theta_{13}=0~(0.2) \,,\label{yupNH}\\
&& y\lesssim 0.030~(0.16)\,~{\rm for~IH~with~} M_1=100\,{\rm GeV}\,~{\rm and~}\sin\theta_{13}=0~(0.2) \,.\label{yupIH}
\end{eqnarray}

\section{Interplay between lepton flavour and lepton number violating observables}

\begin{figure}[t]
\begin{center}
\includegraphics[width=11.5cm,height=7.5cm]{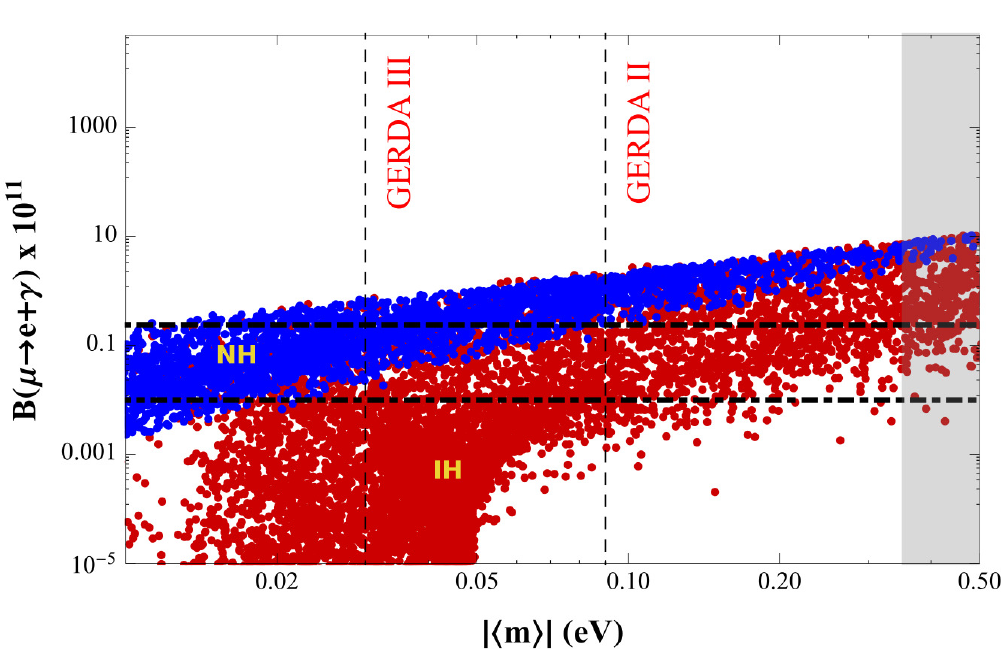}
\caption{$B(\mu\to e+\gamma)$ vs $\meff$ for $M_{1}=100$ GeV and  
$|M_{2}-M_{1}|/M_{1}=10^{-3}$.\label{fig1}}
\end{center}
\end{figure}

Since the flavour structure of the neutrino Yukawa couplings is 
fixed in the present scenarios, correlations among different low energy 
leptonic observables may be a relevant signature of TeV scale type I
see-saw mechanism. 
Indeed, in the simple extension of the Standard Model 
considered, with the addition of two 
heavy RH neutrinos $N_{1}$ and $N_{2}$ at the TeV scale, 
which behave as a pseudo-Dirac particle, 
\emph{a sizable (dominant) contribution of $N_{1}$ and $N_{2}$ 
to the $\betabeta$-decay rate would  imply a ``large'' 
enhancement of the muon radiative decay rate}. 
 In fact, if $\meff \cong |\mefff^{{\rm N}}|$,
where $\mefff^{{\rm N}}$ is given in eq. (\ref{mee3}), it is easy to show that \cite{Ibarra:2011xn}
\begin{equation}
	B(\mu\to e+\gamma)\;\cong\;
	\frac{3\alpha_{\rm em}}{64\pi}\,\left| G(0)-G(M_1^2/M_W^2)\right|^{2}\,\left| 
r \right|^{2}\, \frac{M_{1}^{2}}{M_{a}^{4}}\,
\frac{|\mefff^{{\rm N}}|^2}{z^{2} (f(A))^{2}}\,,	
\label{Bmeg}
\end{equation}
%
where 
$0.5 \lesssim|r|\lesssim 30$ ($0.01 \lesssim|r|\lesssim 5$) for the NH (IH) light neutrino mass spectrum.
 The analytic relation   in eq. (\ref{Bmeg}) 
is confirmed by the results of the numerical 
computation reported in figure~\ref{fig1}, where it is shown
the correlation between the $\mu\to e+\gamma$
branching ratio and the effective Majorana mass in the 
case of sizable couplings between 
the RH (pseudo-Dirac pair) neutrinos and 
charged leptons.
In general, a lower bound on $B(\mu\to e+\gamma)$ within the MEG experiment sensitivity reach
is set for both light neutrino mass  hierarchies (normal and inverted) if
a positive signal is detected by GERDA, $i.e.$ for $\meff\sim 0.1$ eV.

In conclusion, the observation of $\betabeta$-decay in the 
next generation of experiments, under 
preparation at present, 
and of the $\mu\to e+ \gamma$ decay 
in the MEG experiment, could be the 
first indirect evidence for the 
TeV scale type I see-saw mechanism of neutrino 
mass generation.


\end{document}